\documentclass[aps,prl,reprint,superscriptaddress,floatfix,amssymb,amsfonts,showpacs]{revtex4-1}

\usepackage[colorlinks=true,citecolor=blue,urlcolor=blue,linkcolor=black]{hyperref}
\usepackage{graphicx}
\usepackage{graphicx}
\usepackage{dcolumn}
\usepackage{bm}
\usepackage{lipsum}
\usepackage{amsmath}
\usepackage[mathlines]{lineno}

\usepackage{changes}
\begin{document}
\title[]{Direct Observation of Coherent Longitudinal and Shear Acoustic Phonons in TaAs Using Ultrafast X-ray Diffraction}

\author{Min-Cheol Lee}
\thanks{Corresponding author}
\email{mclee@lanl.gov}
\author{N. Sirica}
\affiliation{Center for Integrated Nanotechnologies, Los Alamos National Laboratory, Los Alamos, New Mexico 87545, USA}

\author{S. W. Teitelbaum}
\affiliation{Stanford PULSE Institute, SLAC National Accelerator Laboratory, Menlo Park, California 94025, USA}
\affiliation{Stanford Institute for Materials and Energy Sciences, SLAC National Accelerator Laboratory, Menlo Park, California 94025, USA}

\author{A. Maznev}
\affiliation{Department of 
Chemistry, Massachusetts Institute of Technology, 77 Massachusetts Avenue, Cambridge, MA, 02139, USA}
\affiliation{Institute for Soldier Nanotechnology, Massachusetts Institute of Technology, 500 Technology Square, NE47-598, Cambridge, MA, 02139, USA}

\author{T. Pezeril}
\affiliation{Department of 
Chemistry, Massachusetts Institute of Technology, 77 Massachusetts Avenue, Cambridge, MA, 02139, USA}
\affiliation{Institut de Physique de Rennes, Universit\'e de Rennes 1, UMR CNRS 6251, 35000 Rennes, France}

\author{R. Tutchton}
\affiliation{Center for Integrated Nanotechnologies, Los Alamos National Laboratory, Los Alamos, New Mexico 87545, USA}

\author{V. Krapivin}
\affiliation{Stanford PULSE Institute, SLAC National Accelerator Laboratory, Menlo Park, California 94025, USA}
\affiliation{Stanford Institute for Materials and Energy Sciences, SLAC National Accelerator Laboratory, Menlo Park, California 94025, USA}
\affiliation{Department of Applied Physics, Stanford University, Stanford, California 94305, USA}

\author{G. A. de la Pena}
\affiliation{Stanford PULSE Institute, SLAC National Accelerator Laboratory, Menlo Park, California 94025, USA}
\affiliation{Stanford Institute for Materials and Energy Sciences, SLAC National Accelerator Laboratory, Menlo Park, California 94025, USA}

\author{Y. Huang}
\affiliation{Stanford PULSE Institute, SLAC National Accelerator Laboratory, Menlo Park, California 94025, USA}
\affiliation{Stanford Institute for Materials and Energy Sciences, SLAC National Accelerator Laboratory, Menlo Park, California 94025, USA}
\affiliation{Department of Applied Physics, Stanford University, Stanford, California 94305, USA}

\author{L. X. Zhao}
\author{G. F. Chen}
\author{B. Xu}
\author{R. Yang}
\affiliation{Institute of Physics, Chinese Academy of Sciences, Beijing 100190, China}

\author{J. Shi}
\affiliation{Department of 
Chemistry, Massachusetts Institute of Technology, 77 Massachusetts Avenue, Cambridge, MA, 02139, USA}

\author{J.-X. Zhu}
\affiliation{Center for Integrated Nanotechnologies, Los Alamos National Laboratory, Los Alamos, New Mexico 87545, USA}

\author{D. A. Yarotski}
\affiliation{Center for Integrated Nanotechnologies, Los Alamos National Laboratory, Los Alamos, New Mexico 87545, USA}

\author{X. G. Qiu}
\affiliation{Institute of Physics, Chinese Academy of Sciences, Beijing 100190, China}

\author{K. A. Nelson}
\affiliation{Department of 
Chemistry, Massachusetts Institute of Technology, 77 Massachusetts Avenue, Cambridge, MA, 02139, USA}
\affiliation{Institute for Soldier Nanotechnology, Massachusetts Institute of Technology, 500 Technology Square, NE47-598, Cambridge, MA, 02139, USA}

\author{M. Trigo}
\affiliation{Stanford PULSE Institute, SLAC National Accelerator Laboratory, Menlo Park, California 94025, USA}
\affiliation{Stanford Institute for Materials and Energy Sciences, SLAC National Accelerator Laboratory, Menlo Park, California 94025, USA}

\author{D. A. Reis}
\affiliation{Stanford PULSE Institute, SLAC National Accelerator Laboratory, Menlo Park, California 94025, USA}
\affiliation{Stanford Institute for Materials and Energy Sciences, SLAC National Accelerator Laboratory, Menlo Park, California 94025, USA}
\affiliation{Department of Applied Physics, Stanford University, Stanford, California 94305, USA}
\affiliation{Department of Photon Science, Stanford University, Stanford, California 94305, USA}

\author{R. P. Prasankumar}
\thanks{Corresponding author}
\email{rpprasan@lanl.gov}
\affiliation{Center for Integrated Nanotechnologies, Los Alamos National Laboratory, Los Alamos, New Mexico 87545, USA}

\date{\today}
\begin{abstract}
 Using femtosecond time-resolved X-ray diffraction, we investigated optically excited coherent acoustic phonons in the Weyl semimetal TaAs. The low symmetry of the (112) surface probed in our experiment enables the simultaneous excitation of longitudinal and shear acoustic modes, whose dispersion closely matches our simulations. We observed an asymmetry in the spectral lineshape of the longitudinal mode that is notably absent from the shear mode, suggesting a time-dependent frequency chirp that is likely driven by photoinduced carrier diffusion. We argue on the basis of symmetry that these acoustic deformations can transiently alter the electronic structure near the Weyl points and support this with model calculations. Our study underscores the benefit of using off-axis crystal orientations when optically exciting acoustic deformations in topological semimetals, allowing one to transiently change their crystal and electronic structures.
\end{abstract}

\maketitle
 Ultrafast spectroscopy is a powerful tool for studying structural dynamics in quantum materials through the generation of coherent phonon oscillations driven by femtosecond optical pulses \cite{Dhar1994,Zeiger1992,Thomsen1986,Lee2019}. The coherent excitation of acoustic phonons occurs when an intense femtosecond optical pump pulse launches a strain wave along the surface normal, resulting from thermoelastic effects, piezoelectric effects, or the deformation potential, driven by the photoexcitation of charge carriers  \cite{Thomsen1986,Matsuda2004,Pezeril2007,Pezeril2016,Reis2007,Lee2015}. This strain wave is then measured in the time domain via coherent oscillations in reflection or transmission of an optical probe pulse, when the probe penetrates deeper than the acoustic wavelength. Optical pump-probe spectroscopy has thus been used to observe longitudinal acoustic (LA) phonons with gigahertz (GHz) to terahertz (THz) frequencies in a wide range of materials \cite{Ruello2015}. In contrast, the generation of transverse acoustic (TA) shear modes is less common, as it requires a specific choice of crystal geometry \cite{Pezeril2016,Matsuda2004}. However, the ability to optically excite these acoustic modes at ultrafast timescales could provide a new opportunity for tuning both symmetry and topology in quantum materials, especially in comparison to previous approaches using static strain \cite{Mutch2019} or optical phonons \cite{Weber2018,Sie2019,Luo2021,Aryal2021,Wang2021,Ilan2020}. 
 
 Recently, the advent of ultrafast X-ray free-electron lasers (XFELs) has led to significant advances in the study of structural dynamics, making it possible to directly probe the generation of coherent phonons following optical excitation \cite{Dornes2019,Reis2007,Bostedt2016,Henighan2016}, even at high (THz) frequencies \cite{Henighan2016}. Here, we take advantage of these recent developments in ultrafast X-ray scattering to directly track coherent acoustic phonons in the Weyl semimetal TaAs following femtosecond optical excitation. Time-resolved X-ray diffraction (tr-XRD) was used to probe real-time oscillations in the crystal truncation rod, arising from the structural boundary at the surface \cite{Reis2007,Reis2001,Lindenberg2000}. Due to the distinct orientation of the (112) face, we simultaneously observed both longitudinal and shear acoustic phonon modes. Our data reveals a positive frequency chirp in the longitudinal mode, shown by its asymmetric spectral lineshape, that is absent from the shear mode. This is likely due to phonon softening via ambipolar diffusion of photoinduced carriers. More importantly, our calculations show that both acoustic modes can transiently modify the electronic structure near the Weyl points. These first ultrafast XRD experiments on topological semimetals thus demonstrate the power of this technique for probing and even modifying their properties.
 
\begin{figure}[t!]
	\includegraphics[width=3.4 in]{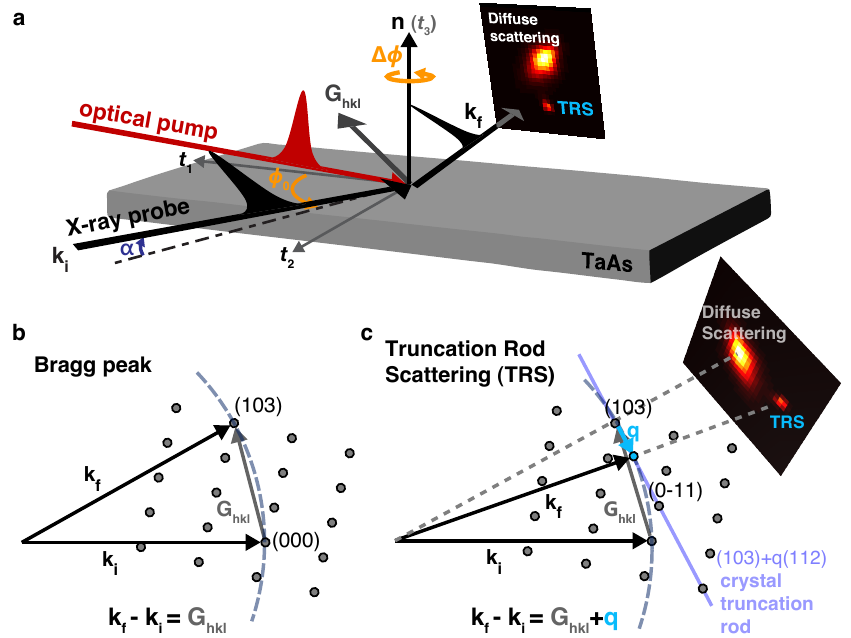}
	\centering
	\caption{Schematic diagram of (a) the time-resolved X-ray diffraction experiment, including the crystal truncation rod scattering (TRS) signal, performed on the (112) face of TaAs, (b) Bragg reflection, and (c) X-ray TRS. The direction of $t_3$ is along the surface normal $n$, while the in-plane coordinates $t_1$ and $t_2$ are along [11$\bar{1}$] and [1$\bar{1}$0], respectively. The incident ($k_i$) and reflected ($k_f$) X-rays as well as the reciprocal vector for Bragg reflection ($G_{hkl}$) are shown with black arrows.}
	\label{FIG1}
\end{figure}

 We performed optical-pump, X-ray-probe measurements at the X-ray pump-probe (XPP) instrument \cite{Reis2007} of the Linac Coherent Light Source (LCLS) XFEL \cite{Bostedt2016} on a 2 mm $\times$ 1 mm, (112) oriented TaAs single crystal, whose growth and characterization is described in \cite{Huang2015}. Fig. 1(a) shows an experimental schematic, where photoinduced changes in TaAs driven by 800 nm (1.55 eV) optical pump pulses (incident fluence = 2.86 mJ/cm\textsuperscript{2}) were probed by monochromatic X-ray pulses at a 120 Hz repetition rate, centered at 9.52 keV. We operated in a reflection geometry with the X-ray probe (optical pump) projected onto the sample at a grazing angle of $\alpha = 3^\circ$ (5.3$^\circ$) in order to match the optical and X-ray penetration depths, allowing us to capture optically-induced lattice modulations \cite{Henighan2016}. Transient changes in the X-ray scattering intensity were measured with an area detector \cite{Blaj2015}. The temporal resolution of our tr-XRD experiments was $\sim$80 fs, with the timing jitter between the optical pump and x-ray probe corrected for on a shot-by-shot basis \cite{Harmand2013}.
 
\begin{figure*}[!t]
	\includegraphics[width=7 in]{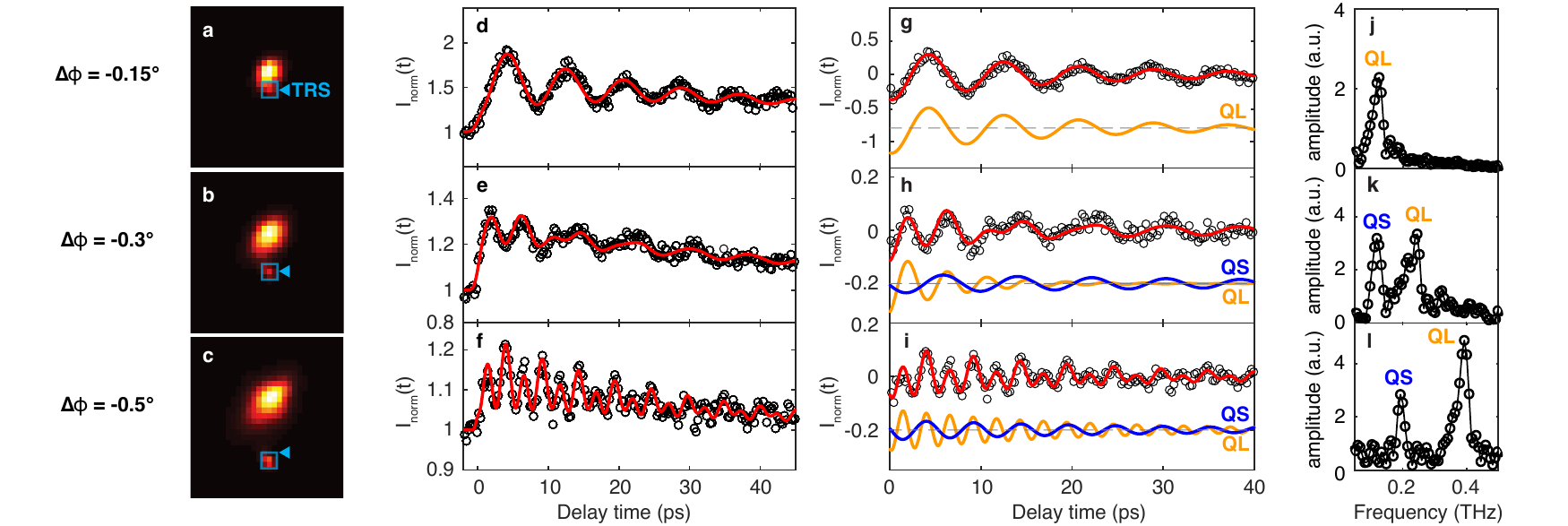}
	\centering
	\caption{(a-c) Two-dimensional XRD images, showing diffuse scattering and TRS signals without pump excitation. Here, variation of the rotation angle, $\Delta\phi$, enables the TRS peak (blue arrow) to be resolved. (d-f) Time-resolved TRS signal traces, integrated over the 3$\times$3 pixel area of the detector ($0.36^\circ$$\times$$0.36^\circ$) indicated by the blue arrow, reveal clear coherent acoustic phonon oscillations. These oscillations were isolated and extracted from (d-f) for $\Delta\phi=$ (g) $-0.15 ^\circ$, (h) $-0.3 ^\circ$, and (i) $-0.5 ^\circ$, along with their spectral amplitudes (j-l). Experimental data (open symbols) was fit (red lines) as described in the text, where the separate fit components for quasi-longitudinal (QL-orange lines) and quasi-shear (QS-blue lines) modes are shown below.}
	\label{FIG2}
\end{figure*}

 The X-ray scattering signal was measured along the (103)-$q$(112) truncation rod, where $q$ defines the wavevector parallel to the (112) surface normal. Detection of the (103) Bragg peak was accomplished by satisfying the Laue condition $\vec{G}_{hkl}=\vec{k}_f-\vec{k}_i$, in which $\vec{G}_{hkl}$ defines the reciprocal lattice vector for the body centered tetragonal ($I4_{1}md$) space group of TaAs \cite{Saini1964, Lv2015,Xu2015}, while $\vec{k}_i$ ($\vec{k}_f$) defines the wavevector for the incident (diffracted) X-ray beam (Fig. 1(b)). Following 800 nm photoexcitation, energy transfer from photoexcited carriers to the lattice occurs within $\sim$10 ps for this (103) Bragg peak (Sec. I in the supplemental material (SM) \cite{supple}).
 
 We then rotated the sample about the surface normal $n$ by a small angle $\Delta\phi$ \cite{surface}. This modifies the scattering condition to $\vec{k}_f-\vec{k}_i=\vec{G}_{hkl}+\vec{q}$, where the wavevector $\vec{q}=q\hat{n}$ is directed along the surface normal, resulting in truncation rod scattering (TRS) (Fig. 1(c)). The incident angle $\alpha$ remains fixed, as is necessary to maintain a constant X-ray penetration depth at grazing incidence. The momentum $q$ can then be determined from  
\begin{equation}
{q}=\Delta\phi{G}_{hkl}k\cos{\alpha}\frac{\sin{\phi_0}\cos{\theta_1}-\cos{\phi_0}\cos{\theta_2}}{G_{hkl}\cos{\theta_3}-k\sin{\alpha}},
\label{eq1}
\end{equation}
 where $k=|k_i|=|k_f|$, $\phi_0$ is the rotation angle that satisfies the Laue condition for the (103) Bragg peak and $\vec{G}_{hkl} = G (\cos{\theta_1},\cos{\theta_2},\cos{\theta_3})$ defines the reciprocal lattice vector in terms of directional cosines with respect to the rotated coordinate frame of the (112) surface $(t_1,t_2,t_3)$ (Fig. 1) (Sec. II in \cite{supple}). 
 
 Fig. 2(a-c) depicts two-dimensional images from our tr-XRD experiments, showing both diffuse scattering and TRS signals. The separation between these two features grows with increasing $\Delta\phi$. Without optical excitation, we can see the bulk diffraction as well as the static TRS signal due to the surface \cite{Robinson1986}. The optical pump pulse then generates a broadband coherent acoustic phonon pulse propagating along the surface normal \cite{Thomsen1986,Henighan2016,Reis2001}. Since the acoustic phonon wavevector $q$ is normal to the surface, x-ray scattering from coherent phonons occurs in the same direction as the static TRS signal, with the magnitude of $q$ given by Eq. (1). By varying $\Delta\phi$, one can then probe different spatial Fourier components of the broadband coherent phonon pulse. 
 
 Fig. 2(d-f) shows transient changes in the TRS signal ($I_{\textrm{norm}}(t)$), obtained by integrating the diffraction intensity over a small $3\times3$ pixel area of the detector (indicated by the blue arrows in Fig. 2(a-c)) and normalizing to the static TRS signal measured prior to pump excitation ($t < 0$). We find that $I_{\textrm{norm}}(t)$ exhibits clear oscillations at sub-terahertz frequencies that scale with increasing $q$ \cite{Reis2001,Henighan2016}. These oscillations result from the interference of the signal scattered from the coherent phonons, which undergoes a phase shift as the phonons propagate deeper into the material, with the static TRS signal \cite{Henighan2016}. In essence, the static TRS signal serves as a local oscillator for heterodyne detection of the coherent phonons. 
 To extract the relevant oscillatory and decay parameters, we fit our data in Fig. 2(d-f) to a function comprised of $f(t)\cdot g(t)$, where the step function $f(t)$ captures the initial rise and $g(t)$ is given by:
 
 \begin{align}
 g(t) = A_le^{-t/\tau_l} + \sum_{i=1,2}A_i e^{-t/\tau_i}\cos(2\pi{f_i}t-\varphi_i).
 \label{gt}
 \end{align}
 The time constants $\tau_l$ ($\tau_i$) represent decays due to lattice cooling (dephasing) times for multiple oscillatory components due to coherent phonons
 . Similarly, $A_l$ represents the amplitude of the slow dynamics, and $A_i$, $f_i$ and $\varphi_i$ denote the respective amplitude, frequency, and initial phase for the oscillations, respectively. 

 In Fig. 2(g-i), we isolate these frequency components by subtracting the effect of lattice cooling (i.e., the first term in Eq. (\ref{gt})) from the overall transients. Fourier transforms of the coherent phonon oscillations clearly show two distinct frequencies (Fig. 2(k-l)). Using Eq. (\ref{gt}) to fit the transient TRS signal for different values of $\Delta\phi$ reveals multiple frequency components. This observation of multiple coherent phonon modes with tr-XRD is surprising, as previous experiments on other materials have predominantly revealed a single LA mode traveling along the surface normal \cite{Reis2001,Lindenberg2000,Henighan2016}. This is due to the fact that shear deformations are not generated under photoexcitation when the surface plane is isotropic, as the transverse lattice displacements that accompany shear wave generation require broken symmetry \cite{Pezeril2016,Matsuda2004,Pezeril2007}.

\begin{figure}[t!]
	\includegraphics[width=3.4 in]{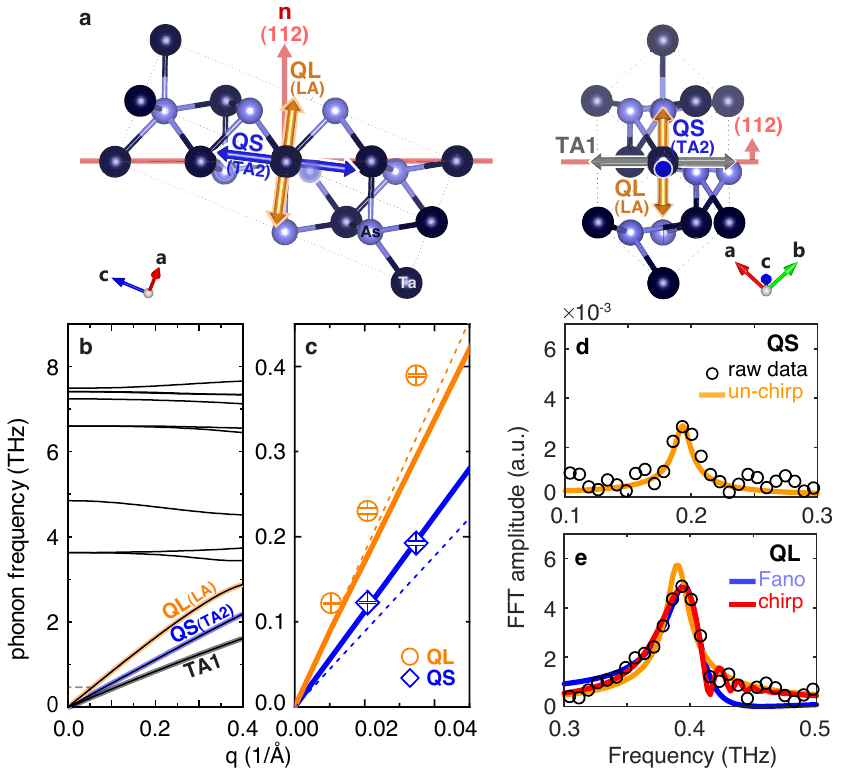}
	\centering
	\caption{(a) Schematic diagram illustrating the polarization of the QL (LA, orange), QS (TA2, blue), and transverse acoustic (TA1, grey) modes with respect to the wavevector along the surface normal ($n$) of the (112) plane, displayed in red. The $a$, $b$, and $c$ axes define the vector norms to the (100), (010), and (001) planes, respectively. (b,c) Calculated phonon dispersion along the (112) direction in TaAs, with the LA (orange), TA2 (blue), and TA1 (grey) modes clearly indicated. (c) Comparison of calculated (lines) and experimentally determined (open symbols) acoustic phonon dispersion measured near the Brillouin zone center (000). Here, the QL (QS) mode is indicated in orange (blue). The solid (dashed) lines are obtained from first-principles (elastic constants) calculations. (d,e) Fits (solid lines) to the spectral lineshape of the (d) QS and (e) QL modes (open circles) for $\Delta\phi = -0.5^\circ$ using Fano resonance (blue), time-invariant (un-chirped frequency, orange), and time-variant (positive frequency chirp, red) frequency models.}
	\label{FIG3}
\end{figure}

\begin{figure}[t!]
	\includegraphics[width=3.4 in]{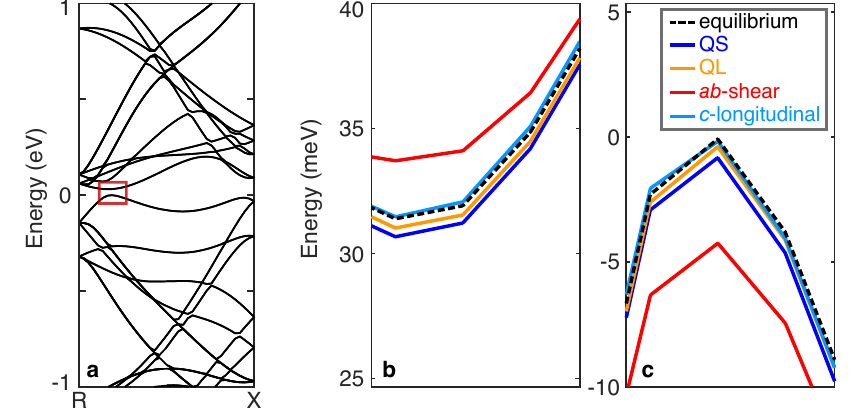}
	\centering
	\caption{(a) Simulated electronic structure of TaAs, obtained from the crystal structures of the 2$\times$2$\times$2 supercell in equilibrium. The red box in (a) indicates the region near one of the Weyl points, magnified in (b) and (c). (b,c) The electronic structure near the Weyl point in (black dashed line) and out of equilibrium after being shifted by 0.2 pm within the supercell for the QL (orange) and QS (blue) acoustic distortions and for $ab$-plane shear (red) and $c$-axis longitudinal distortions (sky blue).}
	\label{FIG4}
\end{figure}

 However, since the (112) surface normal of TaAs deviates from a high symmetry axis (e.g. the $a$ and $c$ axes), the excitation of shear acoustic waves becomes possible \cite{Pezeril2007,Pezeril2016}. Here, photoinduced stress drives both quasi-longitudinal (QL) and quasi-shear (QS) acoustic modes, with displacements that are not purely longitudinal or transverse to the (112) axis. While the generation of shear modes is often forbidden by crystal symmetry, choice of a miscut or reduced symmetry face supports their excitation \cite{Pezeril2007,Pezeril2016,Matsuda2004,Ruello2012,Lejman2014,comment_Dornes,Lemke2018,Juve2020}. 
 
 For insight into the origin of the observed acoustic modes, we used the elastic modulus tensor to calculate the phonon polarization (Sec. III in \cite{supple}). This shows that the atomic vibrations of the LA mode predominately align along the (112) surface normal, with an angular deviation of 7.5$^\circ$ with respect to this axis (Fig. 3(a)). Similarly, the higher-frequency TA mode (TA2) is nearly orthogonal to (112), at $82.5^\circ$ to this direction. In addition, we note that only TA2 was observed in our experiments, as the other mode (TA1) is polarized along the [1$\bar{1}$0] axis. This makes it orthogonal to the sagittal [1$\bar{1}$0] plane and therefore invisible in our TRS experiments.
 
 Furthermore, by using an X-ray probe, we can trace the dispersion of coherent acoustic modes by changing the scattering geometry (Eq. \ref{eq1}) \cite{Lindenberg2000,Henighan2016}, overcoming one limitation of optical experiments \cite{Reis2007}. Recent tr-XRD studies observed shear strain in molecular films and multiferroic oxides \cite{Lemke2018,Juve2020}, but they only focused on Bragg peaks, not the TRS signal, preventing them from measuring the phonon dispersion. We obtained the acoustic phonon dispersion vs. frequency for the QL and QS modes from our experimental data, agreeing well with our elastic tensor and first-principles calculations (Fig. 3(b,c)) (Sec. IV in \cite{supple}). This allows us to conclude that the observed QL and QS modes are the LA and TA2 modes, respectively. 

 Closer examination of the QL mode in Fig. 2(j-l) reveals an asymmetric spectral lineshape, contrasting strongly with the symmetric lineshape of the QS mode (Fig. 3(d,e)). We found that a Fano lineshape failed to capture the observed asymmetry (Fig. 3(e)) \cite{Fano1961}, consistent with the weak electron-phonon coupling for acoustic phonons near the Brillouin zone center \cite{Coulter2019}. Instead, we found that a time-varying frequency given by $f(t) = f_0'(1+Ct)$ reproduces the asymmetric spectral lineshape of the QL mode for $C = 1$ $\textrm{ns}^{-1}$ with minimum deviation from our data (Fig. 3(e)), in contrast with fits based on a time-invariant frequency $f_0$ that explain the QS lineshape (Fig. 3(d)). Furthermore, our model incorporating a positive chirp provides a better fit to the time-domain data (Sec. V in \cite{supple}).
 
 The positive chirp and asymmetric lineshape of the QL mode may be linked to the ambipolar diffusion of photoinduced carriers \cite{Ruzicka2010}. We estimate that these carriers diffuse over a distance of $\sim$70 nm \cite{diff_length}, much larger than the penetration depth of the optical pump (22 nm), but within our X-ray probing depth (500 nm). The resulting carrier density gradient could influence the frequency chirp by changing the charge density distribution and total energy of the system, which would in turn modify interatomic forces and cause phonon softening \cite{Teitelbaum2019}. This would initially reduce the frequency of the QL mode, which would then increase as the acoustic wave propagates deeper into the sample, leading to the observed chirp. Alternatively, the photoinduced carrier density gradient also creates a thermal gradient along the surface normal \cite{Chaban2017}. This can be driven by fast electron-phonon thermalization within $\sim$5 ps, as revealed by our numerical simulations using a two-temperature model (Sec. VI in \cite{supple}). This could soften the QL mode near the surface, which then would harden as the strain wave travels to cooler regions of the crystal. In contrast, the symmetric lineshape of the QS mode may indicate that the atomic potential along the in-plane direction is less sensitive to carrier diffusion or temperature \cite{Timur1977,Klieber2013} than its longitudinal counterpart. 

 The use of an off-axis crystal orientation when optically exciting acoustic deformations also offers a promising route for transiently altering the electronic structure of topological semimetals. Symmetry is closely related to topology, especially in Weyl and Dirac semimetals, where topological phases can be tuned through breaking or restoring electronic and lattice symmetries \cite{Mutch2019,Weber2018,Sie2019,Luo2021,Aryal2021,Wang2021}. Similar to static strain, the dynamic strain resulting from acoustic phonon propagation can alter crystalline symmetry by inducing lattice distortions along a given direction \cite{Ilan2020}. While the generation of LA phonons along [001] does not alter the $4mm$ point group symmetry of TaAs, as the inversion symmetry is already broken within the unstrained lattice, shearing in the $ab$-plane can break mirror symmetries and change the energy and position of Weyl nodes reflected across these planes. Such an in-plane shear is forbidden on a (001) face, but in the (112) oriented crystal used here, longitudinal stress along the normal direction excites both QL and QS modes due to elastic anisotropy in TaAs (Sec. VII in \cite{supple}); importantly, the excitation of a QS mode can occur without shear stress. 
 
 For more insight, we performed first-principles calculations to determine whether the QL and QS lattice distortions can transiently alter the electronic structure of TaAs. We calculated the electronic structure for an undistorted 2$\times$2$\times$2 supercell (Fig. 4(a)) (Sec. VIII in \cite{supple}) to directly compare with snapshots of the non-equilibrium structure following acoustic deformation, and then developed a new theoretical model (Sec. VIII in \cite{supple}) to extract the magnitude of the acoustic displacements from our TR-XRD data. Using these values, we could then model the effect that the QL and QS modes have on the electronic structure by defining a strain wave with an artificial periodicity constrained by the supercell (Sec. IX in \cite{supple}). Importantly, this reveals that dynamical strain from both modes alters the electronic structure near the Weyl points (Fig. 4(b,c)). However, the overall change in their energies and positions is relatively small, and their number is conserved under acoustic distortion.
 
 For comparison, we examined the influence of compressive and shear stress along the high-symmetric [100], [010] and [001] axes. Using the same amount of distortion estimated for the QL and QS modes, we found that $c$-axis longitudinal deformation does not alter the electronic structure with respect to equilibrium, while $ab$-plane shear distortions within the supercell create a noticeable change (Fig. 4(b,c)). This is consistent with our expectation that the Weyl nodes can be modified more effectively by $ab$-plane shearing, suggesting that this is the most significant contribution to changes in the low energy electronic structure of TaAs induced by the QL and QS modes.
 
 In conclusion, we investigated coherent acoustic phonons in TaAs following optical excitation with TR-XRD. Due to the distinct crystallographic orientation, both longitudinal and transverse acoustic modes were excited. We observed an asymmetric spectral lineshape for the QL mode, attributed to phonon frequency chirp arising from the ambipolar diffusion of photoinduced carriers on ultrafast timescales. We also argue using symmetry that the Weyl points in TaAs can be transiently altered by the coherent QL and QS modes, as supported by model calculations. Although this does not represent a true topological phase transition, it does demonstrate that light-driven dynamic strain could induce topological phase transitions in other materials, particularly when the electronic structure depends sensitively on strain\cite{Mutch2019,Weber2018,Sie2019,Luo2021,Aryal2021,Wang2021,Ilan2020}.
 
 \begin{acknowledgements}
 This work was performed at the Center for Integrated Nanotechnologies at Los Alamos National Laboratory (LANL), a U.S. Department of Energy, Office of Basic Energy Sciences user facility, under user proposal 2018BU0083. It was primarily supported through the U.S. Department of Energy, Office of Science, Office of Basic Energy Sciences, Division of Materials Sciences and Engineering via FWP No. 2018LANLBES16 (M.-C. L. and R. P. P.), Contract No. DE-AC02-76SF00515 (S. T., V. K., Y. H., M. T. and D. A. R.) and Contract No. DE-SC0019126 (A. M., T. P., J. S. and K. A. N.). Use of the LCLS is supported by the U.S. Department of Energy, Office of Science, Office of Basic Energy Sciences under Contract No. DE-AC02-76SF00515. J.-X. Z., and D. A. Y. are supported by the Center for Advancement of Topological Semimetals, an Energy Frontier Research Center funded by the U.S. Department of Energy Office of Science, Office of Basic Energy Sciences, through the Ames Laboratory under its Contract No. DE-AC02-07CH11358. N. S. was supported by the LANL LDRD program. This research used resources at the National Energy Research Scientific Computing Center (NERSC), a U.S. Department of Energy Office of Science User Facility operated under Contract No. DE-AC02-05CH11231. We appreciate Matthieu Chollet's assistance in performing experiments at the LCLS XPP beamline. 
\end{acknowledgements}


\begin{references}
\bibitem{surface} The normal vector of the surface (112) is $n = [11t]$ in the real-space coordinates, where $t = 2(a/c)^2$ with $a$ and $c$ of the in-plane and out-of-plane unit cell lengths in the tetragonal symmetry.
\end{references}
\end{document}